\documentclass{ws-jai}
\usepackage[flushleft]{threeparttable}
\usepackage{graphicx}
\usepackage{graphics}
\usepackage{color}
\usepackage{hyperref}
\usepackage{verbatim}

\usepackage{dcolumn}
\usepackage{bm}
\usepackage{lineno}

\graphicspath{{figures/}{figures2/}}
\DeclareGraphicsExtensions{.pdf,.gif,.jpg,.png}

\begin{document}

\newcommand{\dfmux}{DfMux}
\newcommand{\sptthreeg}{SPT-3G}
\newcommand{\oscdemod}{oscillator/demodulator}
\newcommand{\fMUX}{fMUX}
\newcommand{\tMUX}{TDM}
\newcommand{\squid}{SQUID}
\newcommand{\rtHz}{$\sqrt{\mbox{Hz}}$}
\newcommand{\pArtHz}{$\frac{\mathrm{pA}}{\sqrt{\mathrm{Hz}}}$}
\newcommand{\phinot}{\mbox{$\Phi_0$}}
\newcommand{\degree}{\mbox{$^{\circ}$}}
\newcommand{\fortran}{{\tt Fortran~77}}
\newcommand{\CXX}{C++}
\newcommand{\order}{\mbox{${\cal O}$}}
\newcommand{\loopgain}{\mbox{${\cal L}$}}
\newcommand{\const}{\mbox{\sc\small Const}}
\newcommand{\approxlt}{ \stackrel{<}{\sim} }
\newcommand{\approxgt}{ \stackrel{>}{\sim} }
\newcommand{\tauweb}{\tau_\mathrm{web}} 
\newcommand{\tauTES}{\tau_\mathrm{TES}}      
\newcommand{\tauLCR}{\tau_\mathrm{LCR}}   
\newcommand{\Rbolo}{R_\mathrm{bolo}}
\newcommand{\Rnormal}{R_\mathrm{normal}}
\newcommand{\Prad}{P_\mathrm{rad}}
\newcommand{\Pelec}{P_\mathrm{elec}}

\newcommand{\valGammaNE}{0.5}

\newcommand{\mycomment}[1]{ }
\newcommand{\hilight}[1]{#1}

\newcommand{\mynotes}[1]{\newline {\footnotesize {\tt #1 } } }
\newcommand{\revisedtext}[1]{ {\color{blue} #1} }
\newcommand{\note}[1]{\mycomment{#1}}
\newcommand{\mattfixthis}[1]{ {\color{green} #1}}

\newcommand{\partno}[2]{{ #1 (#2)}}

\newcommand\arcdeg{\mbox{$^\circ$}}%
\newcommand\arcmin{\mbox{$^\prime$}}%
\newcommand\arcsec{\mbox{$^{\prime\prime}$}}%

\newcommand\onehalf{\mbox{$\frac{1}{2}$}}%
\newcommand\onethird{\mbox{$\frac{1}{3}$}}%
\newcommand\twothirds{\mbox{$\frac{2}{3}$}}%
\newcommand\onequarter{\mbox{$\frac{1}{4}$}}%
\newcommand\threequarters{\mbox{$\frac{3}{4}$}}%


\title{ICE: a scalable, low-cost FPGA-based telescope signal processing and networking system}

\newcommand{\mcgill}{\dagger}
\newcommand{\wvu}{\ddagger}
\newcommand{\anl}{\star}
\newcommand{\cifar}{\S}
\newcommand{\berkeley}{\diamond}
\newcommand{\chicago}{\ast}
\newcommand{\dunlap}{\circ}

\author{ K.\ Bandura$^{\wvu,\mcgill}$, A.N.\ Bender$^{\mcgill,\anl,\chicago}$, 
J.F.\ Cliche$^\mcgill$, T.\ de~Haan$^{\mcgill,\berkeley}$, M.A.\ Dobbs$^{\mcgill,\cifar}$, \\
A.J.\ Gilbert$^\mcgill$, S.\ Griffin$^\mcgill$, G.\ Hsyu$^\mcgill$, D.\ Ittah$^\mcgill$, 
J.\ Mena~Parra$^\mcgill$, J.\ Montgomery$^\mcgill$, \\
T.\ Pinsonneault-Marotte$^{\mcgill}$, S.\ Siegel$^\mcgill$, G.\ Smecher$^\mcgill$, 
Q.Y.\ Tang$^{\mcgill,\chicago}$, 
K.\ Vanderlinde$^{\dunlap,\mcgill}$,
N.\ Whitehorn$^\berkeley$}

\address{
$^\mcgill$Physics Department, McGill University, Montreal, Quebec H3A~2T8, Canada \\
$^\wvu$LCSEE, West Virginia University, Morgantown, WV 26505 \\
$^\anl$Argonne National Laboratory, High-Energy Physics Division, 9700 S. Cass Avenue, 
Argonne, IL, USA 60439 \\
$^\cifar$Canadian Institute for Advanced Research, Toronto, Ontario M5G~1Z8, Canada \\
$^\berkeley$Department of Physics, University of California, Berkeley, CA 94720, USA\\
$^\dunlap$Dunlap Institute for Astronomy and Astrophysics, University of Toronto, Toronto, Ontario, M5S~3H4, Canada \\
$^\chicago$Kavli Institute for Cosmological Physics, University of Chicago, 
5640 South Ellis Avenue, Chicago, IL 60637\\
}

\maketitle

\corres{$^\S$Send correspondence to J. Mena Parra. E-mail: juan.menaparra@mail.mcgill.ca.}

\begin{history}
\received{(to be inserted by publisher)};
\revised{(to be inserted by publisher)};
\accepted{(to be inserted by publisher)};
\end{history}

\begin{abstract}


We present an overview of the `ICE' hardware and software framework that
implements large arrays of interconnected field programmable gate array
(FPGA)-based data acquisition, signal processing and networking nodes 
\hilight{economically}.
The system was conceived for application to radio, millimeter and sub-millimeter telescope readout systems that have requirements beyond typical off-the-shelf processing systems, such as careful control of interference signals produced by the digital electronics, and clocking 
\hilight{of all elements in the system} from a single precise observatory-derived oscillator.
A new generation of telescopes operating at these frequency bands and designed with a vastly increased emphasis on digital signal processing to support their detector multiplexing technology or high-bandwidth correlators---data rates exceeding a terabyte per second---are becoming common.
The ICE system is built around a custom FPGA motherboard that makes use of an Xilinx Kintex-7 FPGA and ARM-based co-processor. The system is specialized for specific applications through software, firmware, and custom mezzanine daughter boards that interface to the FPGA
through the industry-standard FPGA mezzanine card (FMC) specifications.
For high density applications, the motherboards are packaged in 16-slot crates with
ICE backplanes that implement a low-cost passive full-mesh network between the motherboards in a crate, allow high bandwidth interconnection between crates, and enable data offload to a computer cluster.
A Python-based control software library automatically
detects and operates the hardware in the array.
Examples of specific telescope applications of the ICE framework are presented,
namely the frequency-multiplexed bolometer readout systems used for the
South Pole Telescope (SPT) and Simons Array and the digitizer, F-engine, and networking engine for the Canadian Hydrogen Intensity Mapping Experiment (CHIME) and Hydrogen Intensity and Real-time Analysis eXperiment (HIRAX)
radio interferometers. %

\end{abstract}

\keywords{FPGA, radio interferometer, bolometer readout, correlator} 



\section{Introduction}
\label{sec_introduction}

We present  a general purpose hardware, firmware, and software
framework --- the `ICE' system --- that was developed at the McGill Cosmology
Instrumentation Laboratory
to provide a low cost, large-scale digitization,
signal processing and networking electronics backend for the next generation
of radio to sub-millimeter wavelength telescopes.

In \hilight{designing} the ICE system, we focused on
two specific applications: the digital frequency multiplexing (DfMux)
bolometer readout system for the South Pole Telescope (SPT) and Simons
Array \citep{2014SPIE.9153E..1AB},
and the radio interferometry digitizers,
F-engine, and corner-turn networking for
the Canadian Hydrogen Intensity Mapping Experiment \citep[CHIME,][]{ICE_CORNER_TURN_2016}.

The SPT is a 10-m millimeter-wavelength telescope located at
the Amundsen-Scott South Pole Research Station. It is being upgraded now to
its third generation camera~\citep[\sptthreeg,][]{2014SPIE.9153E..1PB} with \hilight{$\sim$15000}
trichroic, polarization-sensitive Transition Edge Sensor (TES)
bolometric \hilight{detectors, divided between two polarizations and three bands at 95,
150 and 220~GHz. The detectors are read out with the DfMux system}. POLARBEAR2 and the Simons
Array~\citep{2016JLTP..tmp...11S} are 4-m millimeter-wavelength telescopes located
on the Atacama plateau (elevation 5000~m) in Chile. Each telescope will have
\hilight{$\sim$7,500} dichroic, polarization-sensitive TES bolometric
\hilight{detectors, divided between two polarizations and two bands at 95 and 150~GHz.  These are also read
out with the DfMux readout system}. These experiments are optimized to characterize the B-mode
polarization anisotropies of the cosmic microwave background (CMB), a powerful
tool for the detection of weak gravitational lensing and gravitational waves from inflation.

The CHIME telescope consists
of four 100~m~$\times$~20~m fixed cylinders each equipped with 256 dual-polarization
feeds that will continuously image the entire northern sky in order to measure
the distribution of neutral hydrogen in the universe and study the nature of
dark energy \citep{Bandura14,Newburgh14}. \hilight{The telescope will also study pulsars,} Fast Radio Bursts (FRBs), and other transient radio
phenomena using a specialized computer cluster backend.  
The instrument requires electronics that can digitize and
channelize (i.e.,\ divide the 400~MHz data bandwidth into thousands of frequency channels) the 2048 analog
inputs, and \hilight{re-arrange (corner-turn)} the data between the
custom field programmable gate array (FPGA)-based F-engine boards before sending it to the 
\hilight{graphics processing unit (GPU)-based X-engine}. This hardware
must fit within a 2M\$ budget envelope that includes development and
manufacturing.

While these applications have common requirements in terms of high density
digitization, Digital Signal Processing (DSP), and accurate centralized timing,
they also offer very different challenges. Bolometric detectors operated in the
millimeter band typically have sensitivity at the
$10^{-17}$~W/$\sqrt{\mathrm{Hz}}$ level, meaning exceptional care must be
taken to minimize noise or electromagnetic
interference (EMI) produced by the digital backend that can be seen by the detectors.
 These telescopes are
typically located in remote, difficult to access environments such as the
South Pole, the Atacama plateau, and aboard stratospheric balloons.
Maintenance is challenging or impossible at these sites, and remote
configuration and optimization is essential. 
The electronic system must be able to operate in the thin, dry air at high
altitude where convection-based cooling is less efficient than at sea-level.
Meanwhile, large radio interferometers such as CHIME require the digitization
of thousands of inputs at GHz sample rates, followed by the networking and
processing of many Tbit/s of data.

The ICE system is designed to meet a joint set of requirements driven by these challenging applications. These requirements are summarized as follows:
\begin{itemlist}
\item{{\bf High density DSP:}} The system must provide high density programmable DSP, in a compact, rack mounted, power efficient form factor.  Typical applications will use dozens or hundreds of FPGA nodes to acquire and process many thousands of channels in a volume compatible with typical telescope cabins or processing huts. 
\item{\bf High-bandwidth networking:} Radio interferometry requires the processing and networking of a massive data rate
(more than a terabyte per second in the case of CHIME), followed by subsequent data offload to computer clusters using standard protocols such as 10~gigabit Ethernet (GbE).
\item{{\bf One precision clock:}}  To maintain coherence for radio interferometry, and to avoid pickup signals from beating oscillators, the entire system (including switching power supplies, peripherals, Ethernet interfaces, co-processor etc.) must be synchronized with a clock from a single high precision oscillator. The clock and synchronization signals must be distributed to each motherboard and its daughter boards with a low-jitter network that allows for synchronous sampling of data.
\item{{\bf GPS timestamps:}}  All data must be accurately tagged with timestamps coming from a centralized GPS receiver. The hardware must provide a timestamp distribution network \hilight{paralleling} the reference clock. This is particularly important to synchronize data framing across the array, and for pulsar timing and other transient events where timing must be compared across instruments.
\item{{\bf Modular daughter boards:}}  The \hilight{system must be} designed to operate on multiple telescopes and experiments, each requiring specific digitization hardware with large bandwidth and many channels. The nodes must allow the use of \hilight{ off-the-shelf} or custom-made daughter boards to accommodate current and future \hilight{applications}.
\item{\bf{Scalability:}} 
The system must be scalable, such that early prototypes with small channel count can be demonstrated on the benchtop before being scaled to larger sizes.
\item{\bf{Power efficiency:}} The system must offer a high \hilight{operations-per-watt ratio and must allow efficient cooling to reduce cost and increase lifetime.} 
\item{{\bf Low cost:}}  The cost of the system must be affordable within the funding envelope of \hilight{the applications}. 
\item{{\bf Robustness and ease of use:}}  Due to the large number of circuit boards, remote operation, and low maintenance requirements, every key hardware component needs to be `aware'---
capable of identifying themselves on the network, reporting their serial numbers, and detecting existing cable connections between boards. This allows for cabling errors, communication errors, circuit board failures, and other issues to be easily identified. Hardware should be resettable remotely and capable of monitoring temperature, power draw and other diagnostics. The system
should be bootable and configurable over the network to allow for smooth deployment of new software and firmware. 
\end{itemlist}

Developing a new custom platform that can fulfill these requirements is non-trivial
and must make sense in terms of budget, schedule, and risk. Before developing new custom firmware and hardware, we first tried to match off-the-shelf technologies to our requirements. We began using a Xilinx demonstration board with a custom daughter board, and implemented an eight-input correlator using the adaptable and powerful CASPER framework~\citep{Parsons:2009pj} as the core of the firmware.  This system acted as an early prototype of the CHIME telescope correlator. This tested hardware and allowed us to
survey the radio frequency (RF) environment at the CHIME site. The
CHIME firmware continues to use CASPER blocks in a portion of the design. 
\hilight{However, the targeted large-scale projects} made it advantageous to invest the time to
develop dedicated, specialized, and streamlined hardware and firmware \hilight{since off-the-shelf hardware and firmware did not provide the appropriate scaling and
the required engineering}
expertise was available in-house. By redesigning the form factor and using the latest generation of FPGAs, we considerably reduced the
number of circuit boards and the amount of cabling needed to implement the CHIME hardware. The custom architecture allowed us to introduce the clocking and timing network, noise management and remote configuration management features that were required.

The following sections provide a description of the hardware, software, and
firmware which resulted from this engineering effort.
Section~\ref{sec_hardware}
describes the ICE hardware (motherboards and backplanes) and
Section~\ref{sec_software} describes the software and firmware
that efficiently operate all elements of a large hardware array.
Finally, Section~\ref{sec_applications} provides concrete examples showing the
platform in use for the CHIME, Simons Array, and the South
Pole Telescopes mentioned above.


\section{ICE System Hardware}
\label{sec_hardware}

\begin{figure}[htbp]
\centering
\includegraphics[width=.7\textwidth]{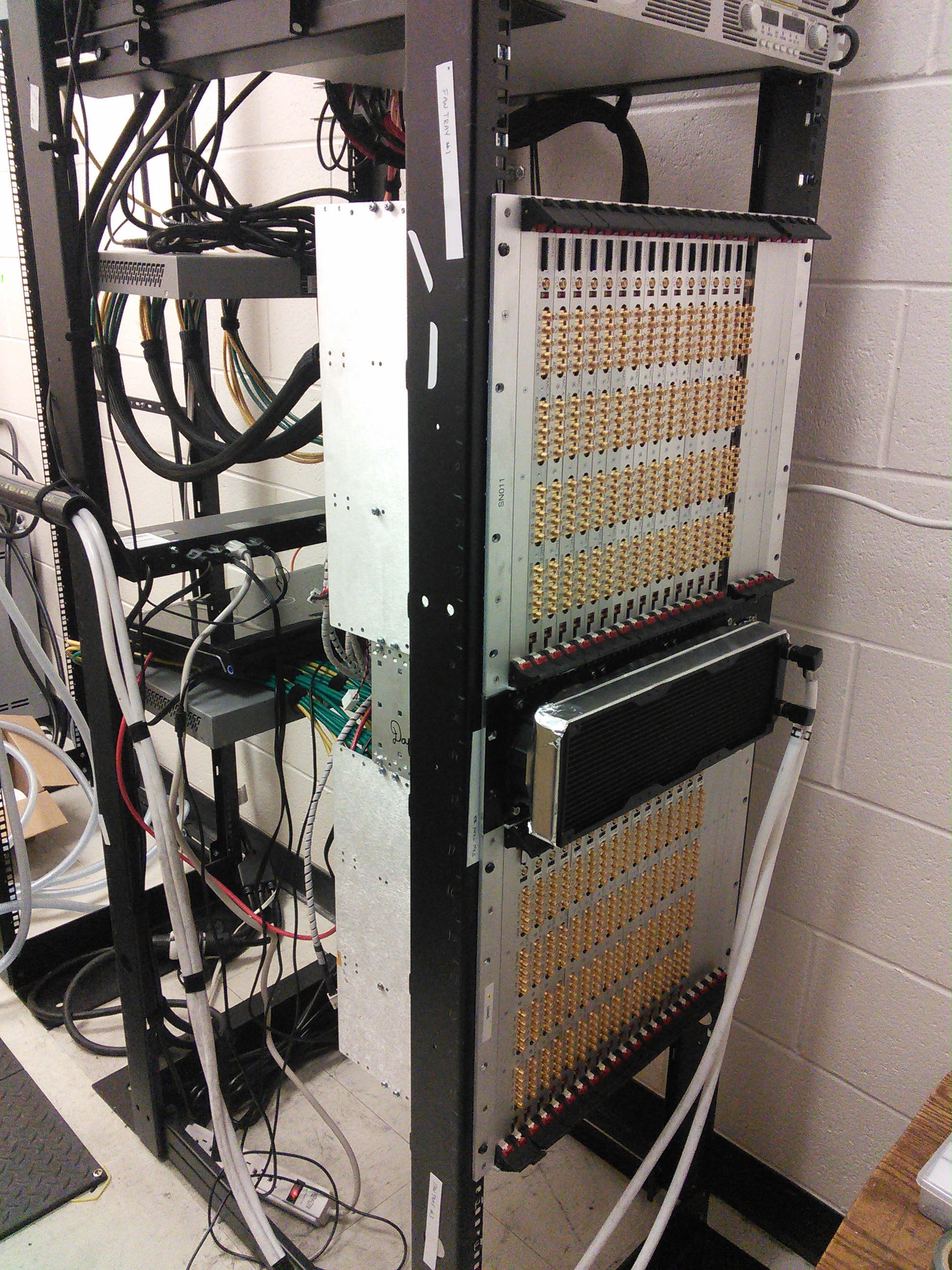}
\caption{Photo showing two fully populated ICE crates mounted in a standard rack.
The motherboards are equipped with the FMC daughter boards designed for the CHIME telescope
(Section~\ref{subsec_chime}). The two crates process 512 analog inputs in total. The fan tray for the
upper crate is installed and equipped with an air-to-water heat exchanger that
cools the air which is circulated through the system.
}
\label{FigBackPlaneFront}
\end{figure}

The ICE framework is composed of FPGA motherboards, application-specific
daughter boards, sixteen-slot crates with custom ICE backplanes, system
software, and application-specific firmware. An example assembly of ICE system
components is shown in Figure~\ref{FigBackPlaneFront}, wherein two crates are
each populated with 16 motherboards. In that example assembly, there are two 8-input, GHz digitizer
daughter boards attached to each motherboard, specializing the system for radio
interferometry applications. The system as shown can digitize 512 inputs and
apply the frequency channelization and corner-turn networking operations. See
Section~\ref{subsec_chime} for more information on this application.

\subsection{FPGA Motherboard}
\label{sec_motherboard}

The ICE motherboard is the primary building block of the system
and features two industry-standard data-acquisition
FPGA Mezzanine Card (FMC) \hilight{slots that connect} to a single Xilinx Kintex-7
FPGA offering twenty-eight 10~Gbit/s high-speed serial transceivers for inter-board networking
and data offload to \hilight{external computers}. An ARM co-processor running Linux
provides 
remote access to the motherboard resources and \hilight{ allows users
to quickly implement high-level algorithms in C or other popular programming languages.}

\begin{figure} [htbp]
\centering
\includegraphics[width=.7\textwidth]{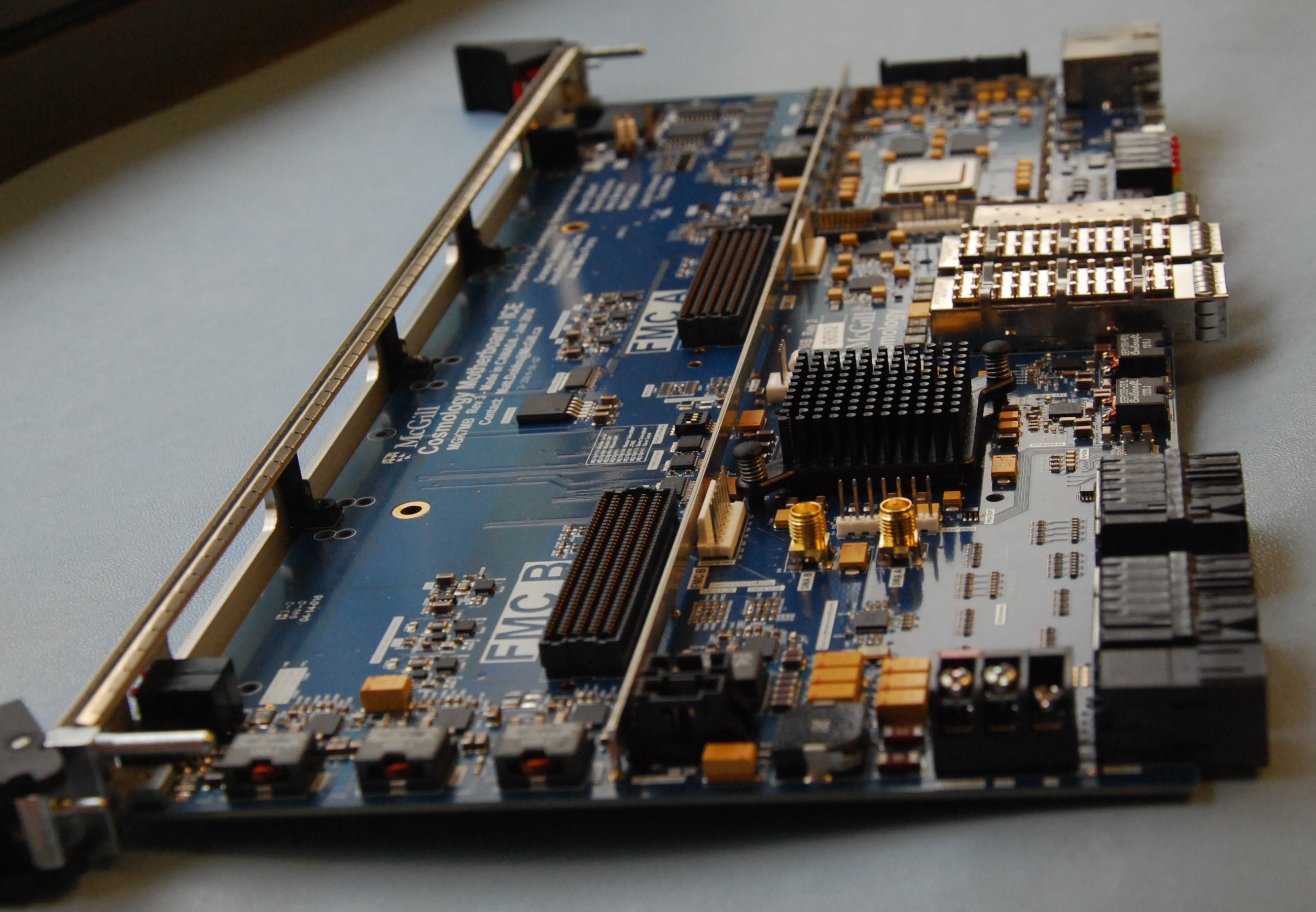}
\caption{Photograph of an ICE motherboard, the FPGA-based processing circuit board that sits at the core of the ICE system. The two industry-standard FMC slots are seen at the left, without daughter boards attached. The FPGA is located below the black heat sink visible near the center of the photo.}
\label{FigICEMBside}
\end{figure}

\begin{figure} [htbp]
\centering
\includegraphics[width=\textwidth]{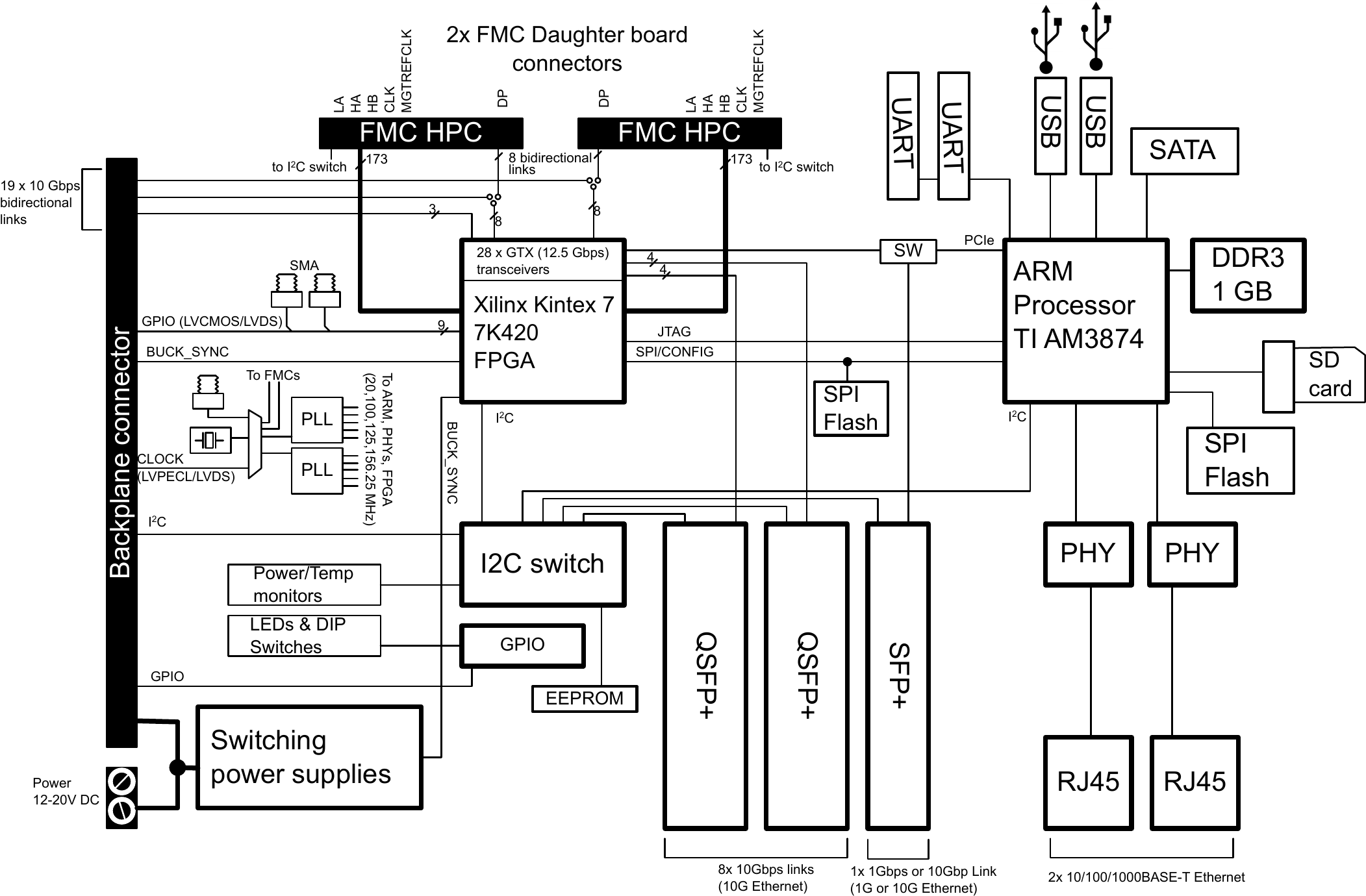}
\caption{Block diagram showing the basic elements of the ICE motherboard.
\hilight{The board is built around a Xilinx Kintex-7 FPGA for real-time
data processing and features two 
slots for industry-standard FMC daughter boards, twenty-eight 10~Gbit/s 
high-speed serial transceivers for
networking, communication and data offload, an ARM co-processor running Linux for 
easy remote access to the motherboard resources and to run high-level applications locally, 
a high precision sub-ps jitter clocking system for timing-sensitive applications, 
and a network of peripherals for board control and
monitoring.}}
\label{FigIceBoardBlockDiagram}
\end{figure}

A photograph of the ICE motherboard is shown in Figure~\ref{FigICEMBside}, and the elements of the motherboard are summarized in Figure~\ref{FigIceBoardBlockDiagram}.
Physically, the ICE motherboard is 37~cm (9U) tall and 16~cm deep. It is tall enough to accommodate two FMC daughter boards and sixteen motherboards in a 9U IEEE standard 1101.10 VME size B crate.
The board is equipped with a latching front panel with handles that allows it to be robustly inserted and secured. The motherboard can be operated as a single stand-alone unit on a bench, using a benchtop enclosure described in Section~\ref{s_two_slot_crate}, or stacked with a 0.8-inch pitch to form high density 16-board crates.

\subsubsection{FPGA}

At the core of the ICE motherboard lies a Xilinx Kintex-7 420T FPGA that has been selected to provide the appropriate amount of
logic, memory and connectivity to efficiently perform the real-time data acquisition, processing and networking needed for the target applications.  The Xilinx-7 series FPGAs have significantly lower power consumption than the previous generation and offer an excellent feature/price ratio.

The FPGA ball grid array package with 901 pads was selected to have enough input/output (I/O) pins to support two daughter boards.
Each pin can capture or generate data at more than 800 MHz using embedded serializers/deserializers and delay lines which are ideal to interface high-speed analog to digital converters (ADCs) and digital to analog converters (DACs) to the slower internal fabric of the FPGA.

On the fabric side, the FPGA offers signal processing resources (block RAM, specialized DSP multipliers) in quantities and ratios that allow efficient implementation of real-time signal processing and the associated data networking for the target applications.
Our typical applications operate on clocks ranging from 200~MHz to 322~MHz. The chip offers sufficient internal routing resources to make all the logic accessible, even in designs that use more than 80\% of the FPGA.

The Kintex-7 420T provides 28 high-speed serial transceivers and is available in speed grades that  can send and receive data at up to 12.5~Gbit/s. We operate these transceivers at 10~Gbit/s and use them advantageously in the following way:
\begin{itemlist}
	\item 15 transceivers communicate with all other FPGAs in the same crate through a direct backplane connection. This \hilight{copper-only link}, from FPGA to FPGA, without active components
	enables \hilight{the} full-mesh high-speed backplane interconnectivity that is essential for radio astronomy applications. For applications that require high-speed serial communications with the daughter boards, backplane transceivers can be redirected to the FMC connectors by changing a few surface mount components on the motherboard. This is useful for high-speed data converters that use fast serial interfaces such as JESD204B.
	\item 4 transceivers from each motherboard pass through the backplane to provide connectivity
	 with other crates through sixteen QSFP+ connectors on the backplane.
	\item 8 transceivers on each motherboard connect to a pair of QSFP+ connectors on the motherboard, typically used to provide eight 10 GbE data offload links. 
	\item 1 transceiver connects through a software controlled switch (labeled `SW' in Figure~\ref{FigIceBoardBlockDiagram}) to either (a) an SFP+ connector to provide an additional 1 GbE or 10 GbE directly from the FPGA, or (b) a high-speed PCIe link to the ARM processor described in Section~\ref{s_arm_processor}.
\end{itemlist}

The FPGA transceivers  use adaptive equalizer techniques that adequately compensate for the losses, reflections, and filtering caused by cables and backplane connections. QSFP+ and SFP+ ports can therefore make use of low-cost passive off-the-shelf copper cables for distances of up to about seven meters. Active copper or optical cables can be substituted for longer distances.

The CHIME telescope uses the fast serial links described above to implement the corner-turn networking that is needed to reorganize the data before sending it to the correlator engine. This application is described in \citet{ICE_CORNER_TURN_2016} and is summarized in Section~\ref{subsec_chime}.

It is worth noting that the FPGA does not have direct access to external memory banks. This design choice is consistent with our target applications, where the FPGA is meant to process data on-the-fly and integrate it in relatively small internal memory before it is processed downstream by a computer system. The FPGA is therefore best suited to acquire, process and network data from a large number of analog inputs, but by design is not intended to implement memory-intensive operations such as large channel-count spatial correlators. \hilight{Since an interface to external memory would require a large number of pins,} the lack of external memory enables us to use a smaller and less expensive FPGA package.

\subsubsection{FMC daughter boards}

The ICE Motherboard is specialized for a particular application by installing
daughter boards that adhere to the industry-standard FMC
specification.
The
motherboard supports both  double- and single-wide FMC cards.
Typically, the daughter boards will be used for \hilight{data conversion}, but
they may also be used for external FPGA memory and other purposes.
\hilight{Custom FMC boards were} designed for the radio interferometry
and multiplexed bolometer readout systems described in
Section~\ref{sec_applications}, but a wide variety of FMC boards are
commercially available and can be used as well. \hilight{Often, the design
process for new applications will involve a proof-of-concept implementation
with commercially available FMCs or vendor demonstration FMCs, after which a custom
daughter board will be built.}

The motherboard accepts FMC daughter boards equipped
with either low- or high-pin count (LPC or HPC) connectors
that connect directly to the FPGA
through 173 single-ended signals, which can also be used as 86
differential clock and data signals. 
\hilight{For daughter boards that require fast serial interfaces,
eight multi-gigabit lanes connecting to the FPGA may also be
rerouted from the backplane connectors to the FMC connectors}.  The 12V,
3.3V and the adjustable I/O voltage 
supplies needed to operate the daughter
boards are provided by the motherboard and can be switched and monitored by
both the FPGA and the ARM processor. The contents of the daughter board EEPROM can
be read to extract the standard Intelligent Platform Management Interface (IPMI)
data which informs the system of the
model and serial number of the board before the main power is applied to it
and before any daughter board-specific FPGA firmware is loaded.

\subsubsection{ARM Processor}
\label{s_arm_processor}

The ICE motherboard is equipped with an ARM processor (Texas Instruments
AM3874) \hilight{that is intended to provide a network-based high-level interface to the motherboards.
Although the FPGA can operate the board without the ARM processor, the latter offers a number of features that make it extremely valuable in managing large arrays of boards.
These include remote programming of
FPGAs, configuration-free networking, and continuous system monitoring. Furthermore, high-level, complex application-specific  functionalities can be efficiently implemented in C, C++ or Python on the ARM processor to free FPGA resources and better isolate the hardware management code from the host-based application.}

The ARM is interfaced to 1~GB of DDR3 memory and two 1 GbE
ports. It can boot from an SD card, on-board flash memory, or using
a network boot protocol. \hilight{The ARM can configure  the FPGA and and 
communicate with it
through a 40~Mbit/s SPI link, the FPGA's JTAG port, and optionally a fast PCIe
link if the FPGA transceiver is not used for an external SFP+ port.}
The ARM processor also provides a number of standard interfaces including
UART, USB, and SATA (see Figure~\ref{FigIceBoardBlockDiagram}), which are
primarily used for debugging in our applications.
Section~\ref{sec_software} describes the software framework that has been
developed to take advantage of the ARM processor.

\subsection{Backplane and Crate}
\label{sec_backplane}

\begin{figure}[htbp]
\centering
\includegraphics[width=0.54\textwidth]{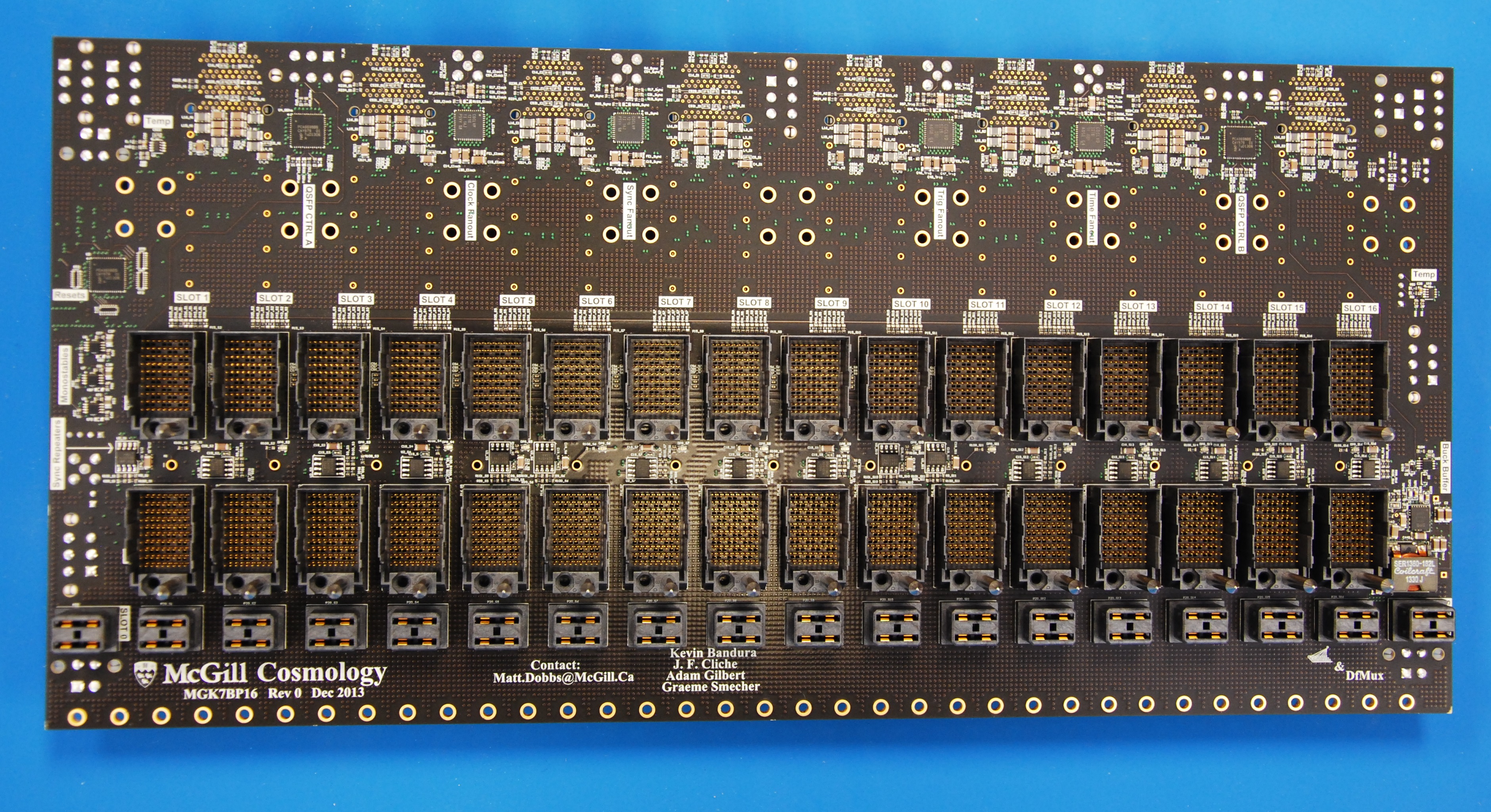}
\includegraphics[width=0.45\textwidth]{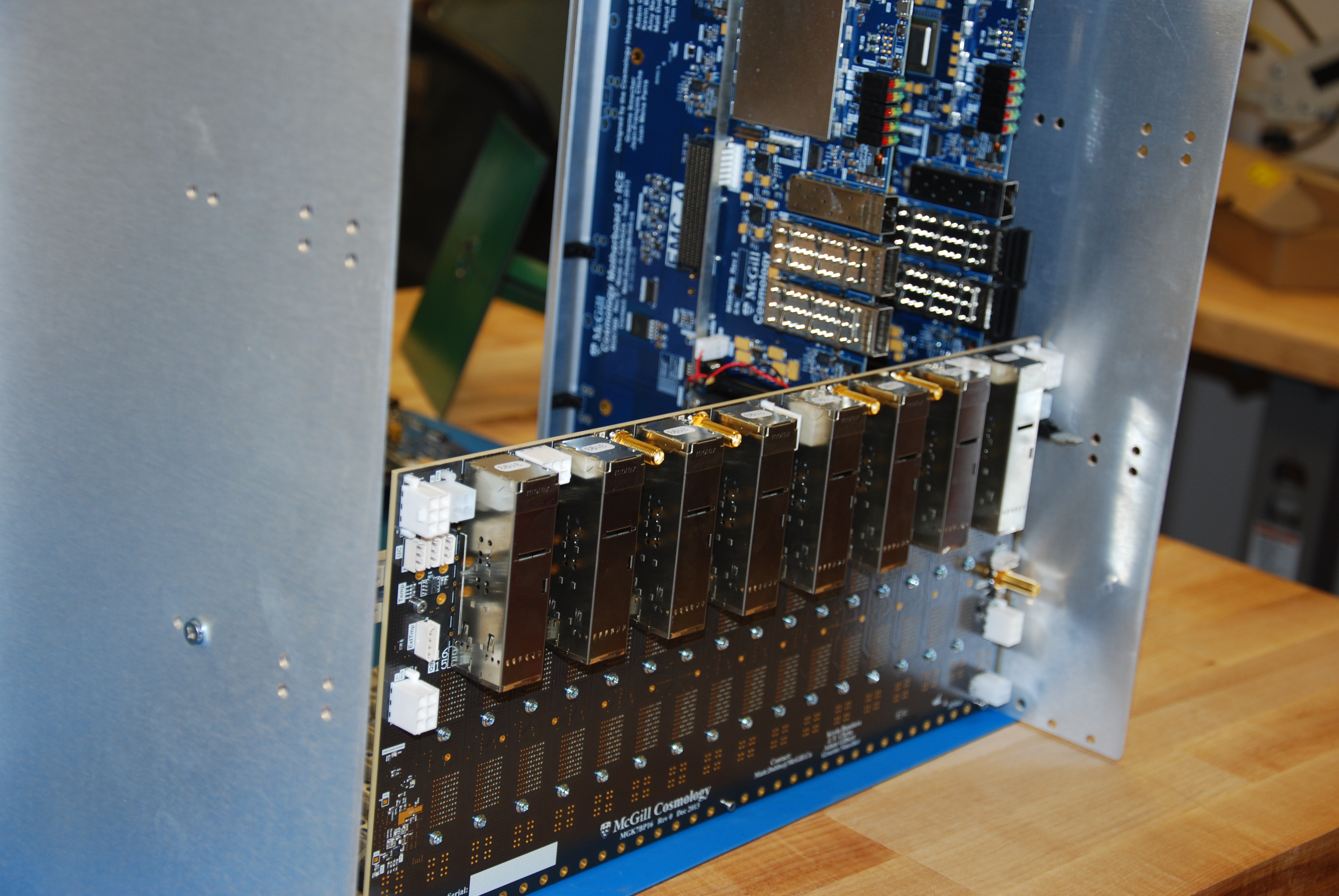}
\caption{(Left) Photo showing the ICE backplane from the
  motherboard mating side. The motherboards
  interface to the backplane is done through high-speed Molex Impact
  Connectors.
  (Right)
  Photo showing the back of the crate-mounted ICE backplane.
  Eight dual-QSFP+ connectors provide data links
  between crates. Power is input on PCIe power connectors.
  Clock, timestamp, and synchronization signals are input on SMA coaxial
  connectors.}
\label{FigBackplanePhoto}
\end{figure}

\begin{figure}[htbp]
\centering
\includegraphics[width=\textwidth]{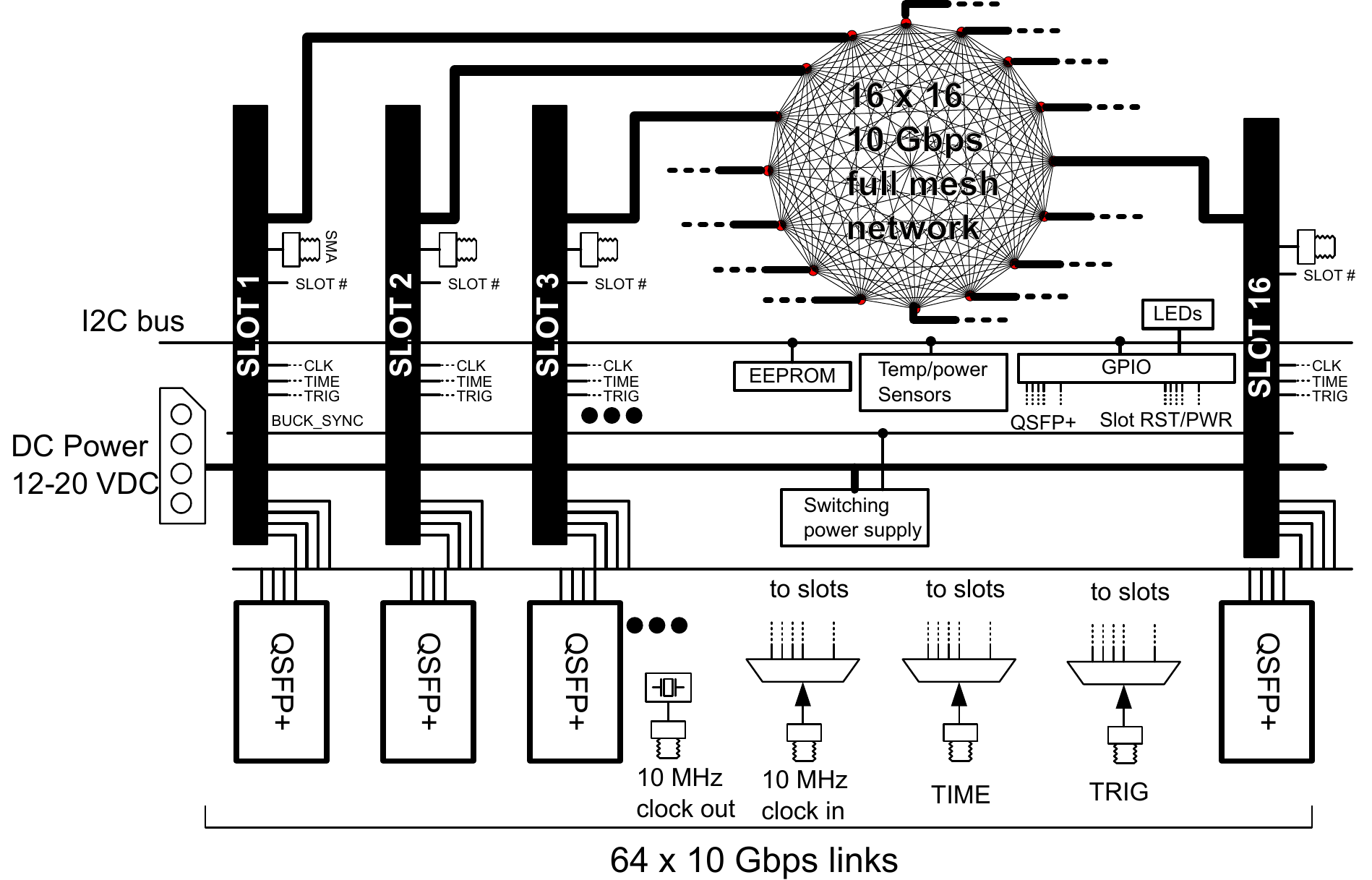}
\caption{Block diagram showing the basic elements of the ICE backplane 
\hilight{that can connect up to 16 ICE motherboards. The backplane distributes power,
a precision clock, GPS timestamp and additional
synchronization and triggering signals to the motherboards.
In addition, it implements a passive 10 Gbit/s full-mesh network for the 16 motherboards in a crate and has sixteen QSFP+ connectors for
multigigabit communications between motherboards in different crates.}}
\label{FigIceBackplaneBlockDiagram}
\end{figure}

For high-density applications, ICE motherboards are packaged in
groups of sixteen into crates equipped with custom ICE backplanes that
 provide power and high-precision timing signals to all its motherboards.
The backplane also provides high-speed, full-mesh connectivity, remote board reset capability,
 and QSFP+ connectors to
interconnect boards between crates. The crates and their backplanes
minimize wiring, increase reliability, and facilitate air or water cooling. A
photo and a block diagram of the backplane are shown in
Figures~\ref{FigBackplanePhoto} and \ref{FigIceBackplaneBlockDiagram}.

The backplane is installed in
9U ($\sim$40~cm)-tall industrial standard Eurocard subrack with
the depth of a VME size B crate (see IEEE standard 1101.10). A crate provides slots for sixteen
ICE motherboards and provides two additional power-only slots that can be used for future expansion such as signal distribution boards.
The backplane interfaces with the lower half of the motherboards only, therefore freeing the
upper portion of the back of the crate to directly expose the
motherboard networking data connectors, switches and indicators.

\subsubsection{High-Speed Networking Fabric}
\label{subsec_full_mesh}

\hilight{
For many applications such as radio interferometry, high-bandwidth
communications between boards in a crate, between crates, and between the ICE
system and a computer cluster are necessary. For example, CHIME has a real time
data rate of about 13 Tbit/s. \hilight{This data 
is rescaled and truncated, then
reorganized
between motherboards by the corner-turn system}, before it is transmitted to the the computer
cluster that does the spatial correlation.

To support these requirements, the backplane implements a full-mesh, passive
network, providing a direct 10 Gbit/s link between each motherboard in a crate
and the fifteen other motherboards. To accomplish this within each link's
attenuation budget, the 24-layer
backplane is constructed of low-loss Megtron 6 material and  interfaces
the motherboards with high-density Impact-series Molex connectors capable of
passing signals up to 25~Gbit/s.  The backplane also has sixteen QSFP+
connectors (capable of 4$\times$10 Gbit/s) intended for communications
between motherboards in different crates. Details of the hardware design are
presented in \citet{ICE_CORNER_TURN_2016}. Operation of the system
confirms that data can be transferred at high rate between all motherboards,
error-free, with a comfortable attenuation margin. }

\subsection{Power System}

\hilight{To make the boards easy to operate in a variety of configurations,}
the ICE \hilight{system} has been designed to operate efficiently
from a single voltage rail supplied with between 13V and 20V DC. An input voltage
as low as 8V can be used if the FMC daughter boards on the motherboards can operate with less
\hilight{than the FMC-standard 12V supply} (as is the case for our
custom FMC boards and the majority of commercial FMCs). An ICE motherboard typically
uses 20W when initially powered, and consumes 60W (excluding the daughter boards)
when running large signal processing firmware that uses the majority
of the FPGA's logic and high-speed transceivers.

\hilight{When used within a crate, the motherboard
power is provided through backplane Molex Impact connectors that can provide up to 10~A per
slot. 
The backplane uses its own on-board buck regulators to generate the voltages needed to power the clock buffers and other active logic. When operated on the benchtop,
power to the motherboard may be provided through screw terminals located directly on the board.}

The motherboard requires a total of nine buck converters to provide the voltage
rails needed by the FPGA, ARM, and peripherals. The frequency and phase of each
switcher can be synchronized independently by the FPGA to a common clock
\hilight{(typically 1 MHz)} derived from
the system reference clock, minimizing the ripple on the supplies and ensuring that
power-supply noise occurs at known and stable frequencies. All voltages are
fixed; smart power controllers are not used as their internal logic operates on
free-running clocks that may introduce hard-to-diagnose interference in our
EMI-sensitive applications.

The FPGA and the ARM can monitor the voltage and current provided by each
switcher. Fault detection circuitry can turn off faulty or overloaded
switchers and raise alarm flags. A block of LEDs indicate the status of each
power rail. An input diode and fuse provide protection against reverse
polarity connections.

\subsection{Cooling} 
\label{sub:cooling}

\hilight{A full crate of ICE motherboards operating at capacity dissipates about 1.5 kW.}
The motherboards are designed to be cooled by forced air. The motherboard cooling requirements are mostly dominated by the FPGA, which is equipped with a low-profile passive heat sink.
In the ICE crates, three 300~CFM fans push air up from the bottom and provide enough airflow across the motherboards to keep the FPGAs below their specified maximum temperature
(85~C). Water-cooled radiators \hilight{
(shown in Figure~\ref{FigBackPlaneFront}) can be used with the fans to cool the air entering the crate, transferring the heat dissipation to water that is easily transported from tight spaces such as telescope cabin RF-shielded rooms.}

A
standard computer heat-sink fan can be installed on
the FPGA heat sink for cooling if the board is used on the benchtop or in the
enclosure described in
Section~\ref{s_two_slot_crate}.

\subsection{Timing System}

Considerable thought was given to the timing system of the ICE system
to optimize flexibility and performance.  For radio interferometry
applications, inter-board clock jitter results in pointing errors on the sky
and must be very carefully controlled. TES bolometers are very sensitive to
EMI and pickup which deposit power in the narrow ($\sim$100 Hz) bandwidth of
the detectors. 
\hilight{Typical commercial systems unfortunately use hardware that uses many free-running oscillators, 
and a common source of EMI for TES detectors arises from
the slowly varying beat frequencies that are created as each oscillator speeds 
or slows, e.g. due to thermal fluctuations.} These considerations motivated the
design of a system with \hilight{a single} clock from which all elements of the
electronics derive their timing, and of a stringent clock jitter requirement
(500~fs for the radio interferometry daughter boards, and 1~ps for radio
interferometry pointing).

The motherboard operates from a single 10 MHz reference that can
be provided from the backplane, the front SMA connector, or an
on-board crystal (which is useful for single-board benchtop applications and
testing). The ARM is aware of which clock is used and can warn users of
improper configurations.

\hilight{The backplane distributes high quality, low-jitter clocking and timing signals to all boards in
the crate. Four single-ended SMA connectors on the backplane allow the user to provide each motherboard with a TTL or
low voltage CMOS (LVCMOS) 10~MHz system clock, a synchronization signal, a trigger signal (to trigger data processing on some external signal), and a time signal (typically GPS timestamps provided in the IRIG-B format). The clock signal
is the most critical, as it ultimately drives the data acquisition system, and uses LVPECL 1:16
buffers with sub-ns rise time and an additive RMS jitter lower than 100~fs.}

All timing and clocking signal tracks \hilight{across the backplanes and motherboards} are impedance controlled and length-matched in order to ensure
good signal integrity and a well-known phase relationship between the signals. The clock lines are differential everywhere 
and terminated appropriately to maintain signal integrity. On the motherboard, the
reference clock is split through a high-speed, low-jitter clock multiplexer
buffer to feed the two FMC daughter boards with a high quality, sub-ps jitter
differential clock needed by some applications to maintain the performance of
high-speed data converters on the FMC daughter boards. The other outputs are fed to two on-board
re-configurable phase-locked loops (PLLs) that generate several different clock
frequencies for use by the ARM (20 MHz, 100 MHz), FPGA processing (10 MHz, 200
MHz), multiple banks of FPGA multigigabit transceivers (125 MHz, 156.25 MHz),
and the Ethernet controllers (25 MHz).

\hilight{Each FPGA can generate programmable clocks and control signals or can read external signals through SMA connectors located on the motherboard or 
micro-coaxial connectors located on the backplane.
}

\subsection{Reset System} 
\label{sub:remote_reset}

\hilight{Individual motherboards can be reset or power-cycled through switches at the back of the boards.} Global manual crate-wide reset and power-cycle switches are also provided on the backplane.

\hilight{For applications that use a large number of motherboards, it is
essential to be able to remotely power down, power-cycle, or reset to recover from a board that is stuck in an invalid software or firmware state, to localize faults, and to disable
problematic circuit boards until maintenance can be
performed.

Consequently, in addition to the manual reset capability, the backplane provides circuitry that allows
any motherboard to remotely power down or reset any other motherboard in the crate, either individually or
crate-wide, with commands sent over the backplane's I2C bus.

}

\subsection{Automatic Hardware Identification System} 
\label{sub:hardware_id}

\hilight{Every motherboard, backplane, and FMC daughter board is equipped with an EEPROM that holds industry-standard IPMI metadata describing the board model, revision, serial number, manufacturer, etc. The backplane also allows each motherboard to query which slot it is plugged into. When the array is booted, the ARM processors gather all the data and advertise the presence of the hardware on the network to allow the control software to quickly build or validate the interconnections of hardware in the ICE system.}

\hilight{Information (technology, length, voltages, manufacturer, etc.) associated with cables plugged into the motherboard and backplane QSFP+ and SFP+ ports is stored in EEPROMs incorporated in the industry-standard cables} and can also be accessed by the ARM processor or the FPGA. The connectivity of the array can automatically be inferred by matching the information read at both ends of cables.

\hilight{This `awareness' of the  ICE system hardware allows a full system configuration report and automatic detection and documentation of system hardware changes.}

\subsection{Peripherals}

The ICE motherboard and backplanes are equipped with I2C control and monitoring networks that are accessible to both the FPGA and the ARM processor. Any motherboard can access the backplane resources by using the multi-master mode allowed by the I2C protocol.

This bus allows the software to:
\begin{itemlist}
\item \hilight{Control peripherals such as LEDs and read the states of DIP switches}.
\item Control and assess the state of the power systems on both the motherboard and backplanes.
\item Monitor the voltage and current on each power rail.
\item Monitor temperature sensors throughout the system (including the FPGA die temperature sensor and external sensors connected to the backplane).
\item Connect to an external fan controller so the crate fan speed can be statically or dynamically controlled.
\end{itemlist}

\hilight{In addition, each motherboard and backplane is also equipped with a `heartbeat' LED that connects directly to the
corresponding FPGA so the state of the system can be assessed at a glance by the users.}


\subsection{Two-slot crate for benchtop applications}
\label{s_two_slot_crate}

\begin{figure}[!htbp]
  \centering
    \raisebox{-0.5\height}{\includegraphics[width=0.48\textwidth]{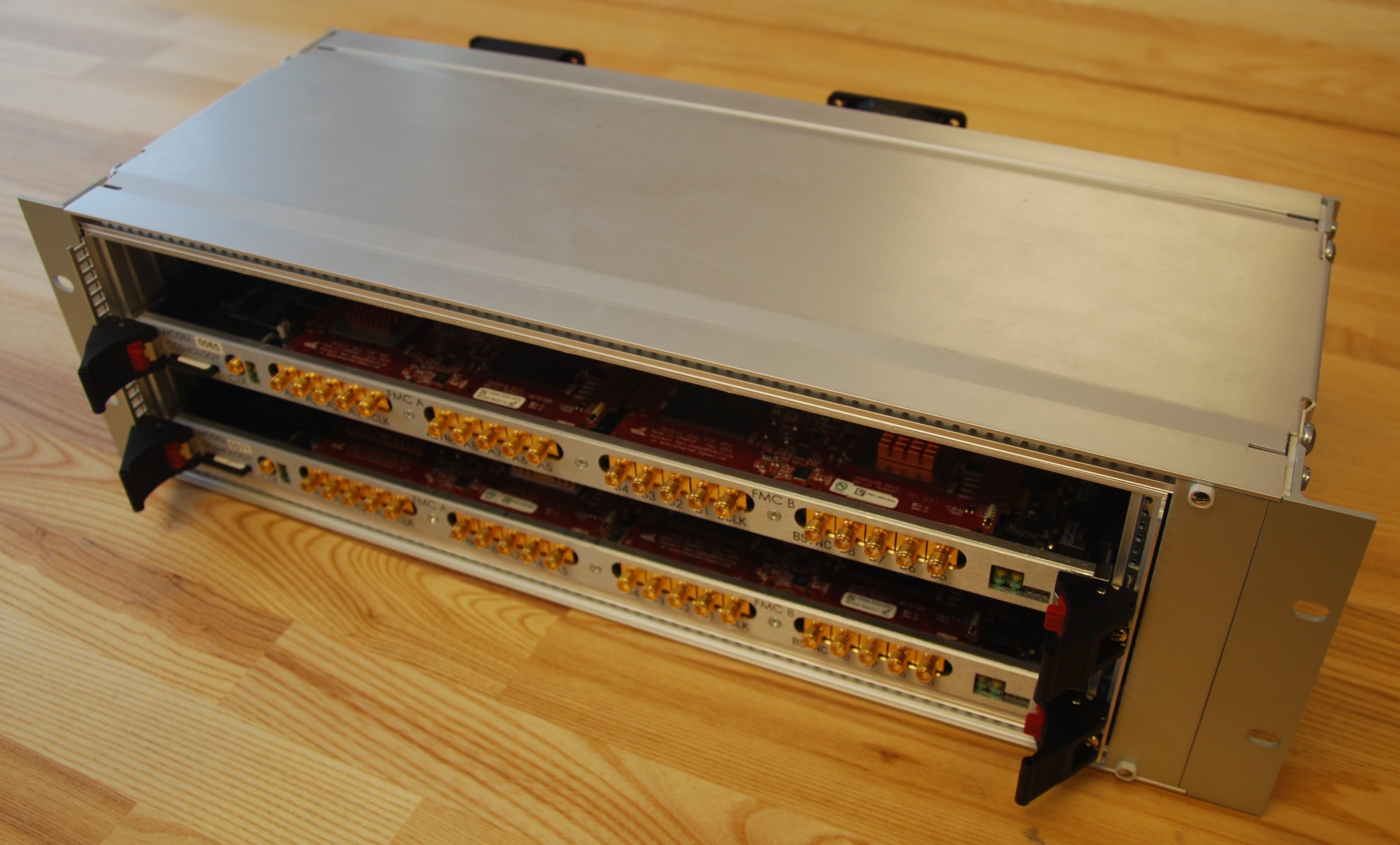}}
    \raisebox{-0.5\height}{\includegraphics[width=0.48\textwidth]{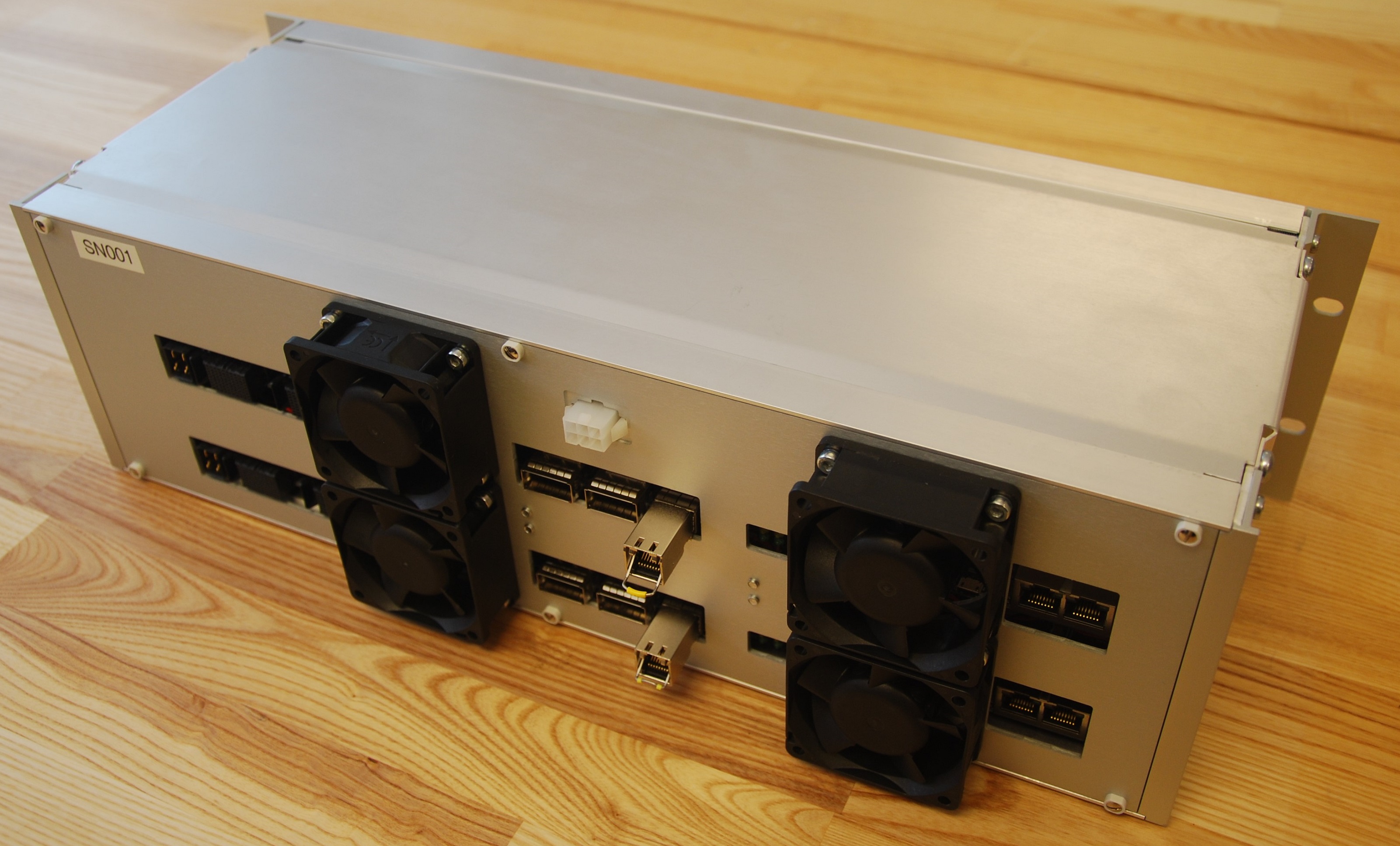}}
    \caption{Small ICE motherboard enclosure for benchtop or small applications. The enclosure is rack-mountable and 3U tall, supporting up to 2 ICE motherboards.
    A custom rear panel exposes the I/O ports of the ICE motherboards.
    }
  \label{FigEnclosure}
\end{figure}

\hilight{
To facilitate the operation of ICE motherboards in small systems for
prototyping and for board testing, a small enclosure intended for one or two
ICE motherboards has been designed (Figure~\ref{FigEnclosure}). This enclosure supports,
cools, and protects the boards. It can be operated on the
benchtop or mounted in 3U of rack space. It uses
small 60~mm fans mounted at the rear which, in conjunction with a 50~mm fan
mounted directly to the FPGA heat sink, are adequate for cooling the boards.
These fans are also much quieter than the ones used in the full ICE crates,
making it more comfortable for use in close proximity to developers.
}

Power is supplied to the enclosure via a 6-pin Molex connector; inside the
crate there is a terminal block that distributes power to the fans and ICE
motherboards. The rear panel exposes the ICE board's primary I/O ports and has
been designed such that in the future, a 2-slot backplane could be used in
situations where a full crate is not required but direct links between two ICE
motherboards over 10 Gbit/s links are desirable.

\section{ICE firmware and software}
\label{sec_software}

The ICE framework relies on three pieces of software or
firmware: (1) the \hilight{Python-based} control software running on an external control
computer to boot, survey, initialize and monitor the array system while executing
system-wide functions,  (2) the ARM processor firmware that is used to
interpret and execute commands sent over the network from the control computer or user, 
and execute high
level asynchronous algorithms, and (3) the FPGA firmware that is typically
responsible for fast, complex, application specific real-time signal processing
\hilight{and for interfacing with the daughter boards.}

The ICE framework includes a core library that provides a unified architecture and
interface for each of these elements. Application-specific firmware is then
built on this core library, which is described briefly in the following
sections.

\subsection{Python-based System Control Software}
\label{subsec_icecore_sw}

The control software provides the tools necessary to configure, operate, and
monitor the hardware \hilight{across the array} in an application-specific
manner.  \hilight{It is built upon} the Python package (IceCore) which is part
of the ICE framework library and provides a basic set of services common to
the ICE hardware.
\hilight{Through the IceCore package,  telescope control systems can
make extensive use of the hardware's self-awareness and a seamless interface
to the ARM processors located on each motherboard to exploit the parallel
computational power of the array and offload tasks from the control computer.}

\hilight{The first step in creating an application is to define Python objects that,
through their attributes and methods, represent each hardware element and its interface.}
The IceCore package provides basic
objects that can be extended to the specific application. For instance, an FMC
daughter board will inherit the generic IceCore \texttt{Mezzanine} object and
will add sub-objects such as ADCs, PLLs, monitoring devices, etc.\ and the
interface to access and operate those resources.

The IceCore package provides an \texttt{IceBoard} object that
represents the ICE motherboard. Using a JSON\footnote{\url{www.json.org}}
 Remote Procedure Call (RPC)
protocol~\citep{Tuber_EWiLi_2012}, it transparently exposes all the functions
that are available on the ARM processor as if they were local methods without the need
to explicitly code interactions at the Python and network layers.
\hilight{This allows dynamic update and extension of the functionalities offered by IceCore library purely through firmware updates on the boards, with no corresponding changes to the control software.}
A basic \texttt{IceBoard} object provides methods to monitor system
voltages and temperature, probe the peripherals, and program the FPGA. The
application typically extends the \texttt{IceBoard} object to add the
functions that are available in the specific FPGA firmware.
Similarly, an \texttt{IceCrate} object is provided to represent the backplane hardware. Users can also
create any arbitrary object that represents other hardware linked to the experiment.

The list of all the objects needed in an experiment and their relationship is stored in a structure
called a \emph{hardware map}. In IceCore, the hardware map is implemented as a Structured Query Language (SQL) relational
database where each \texttt{IceCrate} object is linked to sixteen \texttt{IceBoard} objects, which in turn are linked to
two \texttt{Mezzanine} objects, and so on. The Python SQLAlchemy\footnote{\url{www.sqlalchemy.org}} package is used to hide the complexity of the
database engine such that the hardware map can be used intuitively as a list of objects by the user.
For complex applications, however, the power of the SQL engine can still be used to select
subsets of elements that meet specific connectivity or attribute criteria.

The hardware map of an experiment can be specified in a YAML\footnote{\url{www.yaml.org}} configuration file with various levels of
detail. \hilight{The} \texttt{IceCrate}, \texttt{IceBoard}, \texttt{Mezzanine}, and other user elements and their
relationships can be specified explicitly in this file. In this case, all
the specified objects will be instantiated when the application is loaded, and the system is ready to
run. The software can optionally probe to confirm that the hardware with the right
model and serial number are in the expected locations.

For
systems with a large number of circuit boards, keeping track of the system
configuration is a daunting and time consuming task prone to user error.
\hilight{To solve this problem}, the ICE software can be used to probe
the model and serial number of each hardware element of the ICE
system.
\hilight{
It can also also use the multicast DNS (mDNS) protocol
implemented on the ARM to obtain the IP address of each motherboard 
on the local network along with the information gathered from its 
motherboard, backplane, and daughter-boards.}
The control software can therefore use the IceCore package to discover the hardware elements that correspond to specified selection criteria and automatically build the hardware map. With this functionality, the
system can self-document its hardware configuration, capturing changes such as
circuit board swaps and even which SFP+ or QSFP+ cable is plugged into each
connector. This eliminates the need to maintain exhaustive lists of hardware and IP addresses. 
\hilight{Automatic discovery and mapping} also speeds the process of swapping in spares and tracking
down connectivity errors, greatly improving the efficiency of telescope
operators and hence the observing efficiency of the instrument.

In order to allow efficient operation of large arrays of ICE motherboards and other related objects,
\hilight{IceCore allows access to attributes and methods on a collection of hardware objects
in the same manner as for a single object.
Collections of objects are created using hardware map queries}.
For example, \hilight{
\texttt{myBoards=myHardwareMap.query(IceBoard)} returns a collection of all the IceBoard
objects in the current hardware map. In this case,
\texttt{myBoards.UploadBitstream(’filename’)} would
concurrently program the FPGA on every board in that collection.} Since queries are direct Python representations of SQL queries, they have the power to join,
filter, and subquery.
This powerful tool can be used to perform operations on any subset of the boards meeting a particular \hilight{criteria, such as} crate number, slot number, etc.

\hilight{When accessing collections of objects,} IceCore makes use of the Python Tornado\footnote{\url{www.tornadoweb.org}} package  to run concurrent
method calls asynchronously, so the user does not have to wait for a method to finish before starting
another \hilight{or write complex process threading code}. 
This means that time consuming operations, \hilight{such as accessing slow devices on the motherboards} or acquiring large amounts of data, do not
prevent other requests from starting and finishing.
Any transaction between a control computer and the ARM through HTTP can be run asynchronously.
This can significantly increase the
speed of an application that is operating on large arrays.
In the above example, this means the time required to upload a bitstream and configure the FPGA
takes 20 seconds whether there is one or hundreds of FPGAs in the array. Users can
use the Tornado library to make their own asynchronously called methods to improve parallelism.

\subsection{ARM Firmware}
\label{subsec_arm_fw}

\hilight{The ARM firmware has three main roles: to provide a network-based remote interface for control and monitoring of the ICE motherboard and its surrounding hardware, to allow programming and communication with the FPGA, and to locally execute high level, non-real-time asynchronous algorithms.}

During bootup, the program code on the ROM within the ARM processor loads a stripped-down version of the Universal Boot Loader (U-boot\footnote{\url{www.denx.de/wiki/U-Boot}}) bootloader from
the ICE motherboard embedded Flash memory or from the SD card into its internal RAM. This minimal U-boot configures the DDR3 SDRAM memory and loads a full version of U-Boot into it.
In turn, U-boot loads from the DDR3 memory and starts the Linux kernel
equipped with all the drivers required to access the ICE motherboard's hardware.  The control is then
passed to the kernel, along with the location of the filesystem that contains all the applications needed to provide the core services. The full U-boot build, the Linux kernel, and the root filesystem may be located on the local non-volatile storage (on-board Flash memory, removable SD card) or transferred from the control computer over the network.
Once Linux has booted, the ARM obtains an IP address via DHCP, \hilight{advertises its presence and its surrounding hardware on the network using the mDNS protocol,} and starts the
application stack found on the root filesystem. 

The web server is the main access point \hilight{for commanding and controlling}
the ICE motherboards over the network.
\hilight{Access to the ARM functions (such as operation of the board hardware)} is provided through HTTP requests containing JSON
strings that describe the name of the function to execute (e.g.,
\verb+get_motherboard_temperature+) and its parameters. An application server
parses these requests, calls the corresponding functions in its C-language
libraries, and repackages the results into JSON strings in an HTTP response.
This \hilight{JSON RPC} interface allows easy and standardized access to the board's functions from any programming language. The ICE framework comes with a core ARM library that provides the basic
functions needed to monitor and control the board's hardware and to program
and establish a basic communication link with the FPGA. Application-specific
libraries can be added to extend support for specific FMC daughter boards and FPGA
firmware, if needed.

The RPC interface can be used to offload and parallelize tasks that would be
taxing for the control computer, and to isolate the hardware-specific  code
(such as accessing internal FPGA registers) from the higher-level application
code running on the control computer.

The web server also serves a basic set of web content (HTML, CSS, JavaScript, PDFs,
etc.) allowing users to visualize, and external control software to monitor, the state of the ICE system (temperature, power) and providing basic
controls. Dynamic content and controls within these pages use Javascript to access the board functions over the same JSON-RPC
interface described above. The web pages also provide access to the ICE system's latest documentation.

\subsection{FPGA Firmware}

The firmware running on the FPGA is mostly application and daughter board specific. A library of basic
FPGA physical constraints (i.e., pinout) and core building blocks are maintained as a shared resource to help build
applications. The core blocks include a memory-mapped serial interface with the ARM processor, basic clock
module, switcher synchronization, and an IRIG-B time encoder/decoder.
These core blocks are built into a small application example that specialized firmware
is built on top of.

The FPGA firmware is typically developed using Xilinx's Vivado toolset.
Compiling a large application
that uses most of the FPGA fabric currently takes about two hours.  The resulting bitstream can be stored on the local
non-volatile memory of the ICE motherboard to be programmed automatically when the system boots or 
it can be sent over the network by the control software. Motherboards in an array can operate
different firmware if needed, while maintaining interoperability for core functionality through the ICE control software.

\section{Applications}
\label{sec_applications}

The ICE system has been broadly adopted for the readout of TES bolometers for observations of the CMB 
and high-bandwidth radio interferometers being built for observations at hundreds of megahertz.
More details on these implementations are provided below to illustrate the capability of the ICE platform.


\subsection {Multiplexed Bolometer Readout: Cosmic Microwave Background Experiments}  
\label{subsec_dfmux}

Measurements of the CMB temperature and
polarization anisotropies are at the foundation of  the
Lambda Cold Dark Matter ($\Lambda$CDM) cosmological model~\citep[e.g.,\
][]{Hinshaw:2012aka,Sievers:2013ica,Ade:2013zuv,Ade:2014afa,Hou:2012xq}.
Presently, millimeter-wavelength experiments are racing towards high-sensitivity
measurements of the B-mode component of CMB polarization  \citep[see, e.g.\
][]{Kamionkowski:2015yta}. At small angular scales, gravitational lensing
dominates the B-mode signal and provides information about the mass
distribution of the Universe. At large angular scales, gravitational
waves produced during inflation imprint a characteristic signal in the CMB B-mode polarization
\hilight{which is perhaps the only accessible observation today that could provide a direct
probe of inflationary theories.}

For ground-based and balloon-borne telescopes, TES
bolometers have been the detectors of choice for more than a decade.
TES detectors are limited primarily by uncorrelated noise sources such as photon shot noise,
so focal planes employing many thousands of detectors are being built to
reach  the sensitivity necessary to observe the faint B-mode
polarization. The readout of \hilight{large sub-Kelvin TES arrays} is a key
technological challenge. The limited cooling power of the sub-Kelvin cryogenic
focal plane dictates that only a limited number of wires can traverse the heat
gap between the various cold stages. The ICE system, customized for the
\dfmux\
readout system for TES detectors, is being commissioned now for deployment on
several third generation CMB polarization experiments including
\sptthreeg~\citep[$\sim$15,000 TES detectors]{2014SPIE.9153E..1PB},
POLARBEAR2~\citep[$\sim$7,500 TES detectors]{2014JPSCP...1a3108M} and the
Simons array~\citep[two additional telescopes extending POLARBEAR2, each
with $\sim$7,500 TES detectors]{2016JLTP..tmp...11S}, 
introduced in Section~\ref{sec_introduction}.

The ICE-system \dfmux\ is  based on a previous generation of FPGA-based
readout~\citep{2008ITNS...55...21D} that was used for the SPT
polarimeter \citep[SPTpol, ][]{2012SPIE.8452E..1EA},
POLARBEAR~\citep{2012SPIE.8452E..1CK}, and the balloon-borne E and B
experiment \citep[EBEX][]{doi:10.1117/12.857596}. This earlier system provided
a multiplexing factor of up to 16 detectors per pair of wires. The ICE-system
\dfmux\ electronics are presently capable of multiplexing up to 128 detectors
per pair of wires.
The achievable multiplexing factor is defined by both the
\dfmux\ warm readout electronics and the cryogenic multiplexing circuits.
Presently \sptthreeg\ and POLARBEAR2/Simons Array are configured for
multiplexing factors of 68 and 40, respectively.

The \dfmux\ readout system is described in \citet{2014SPIE.9153E..1AB},
briefly: Each TES bolometer (located on the 250~mK cold stage) is part of a
series-resonant LC circuit tuned to a unique frequency. The voltage biases
for all bolometers in a readout module are applied as a
frequency-comb of sinusoids injected by the room temperature electronics to the focal
plane. The current flowing through the bolometer comb is measured by a
Superconducting Quantum Interference Device (SQUID) preamplifier located on
the 4~K fridge stage. Through use of tuned circuits, the bandwidth of the
bolometer’s Johnson noise, which would otherwise contribute to the noise in
all other channels of the module, is reduced. To ensure that the highly
nonlinear SQUID operates in its most linear region, the current flow into the
SQUID is nulled using a feedback system known as Digital Active
Nulling \citep[DAN,][]{2012SPIE.8452E..0ED}. This nulling signal is injected
by the warm electronics at the SQUID input to cancel the bias sinusoids and
sky signal. The comb of error signals from the SQUID output are digitized and
then transmitted to a bank of demodulators implemented with firmware in the
ICE motherboard FPGA, which mix the signals back down to base-band. These
signals are used to adjust the nuller signals appropriately. 

\begin{figure} [htbp]
\centering
\includegraphics[width=.7\textwidth]{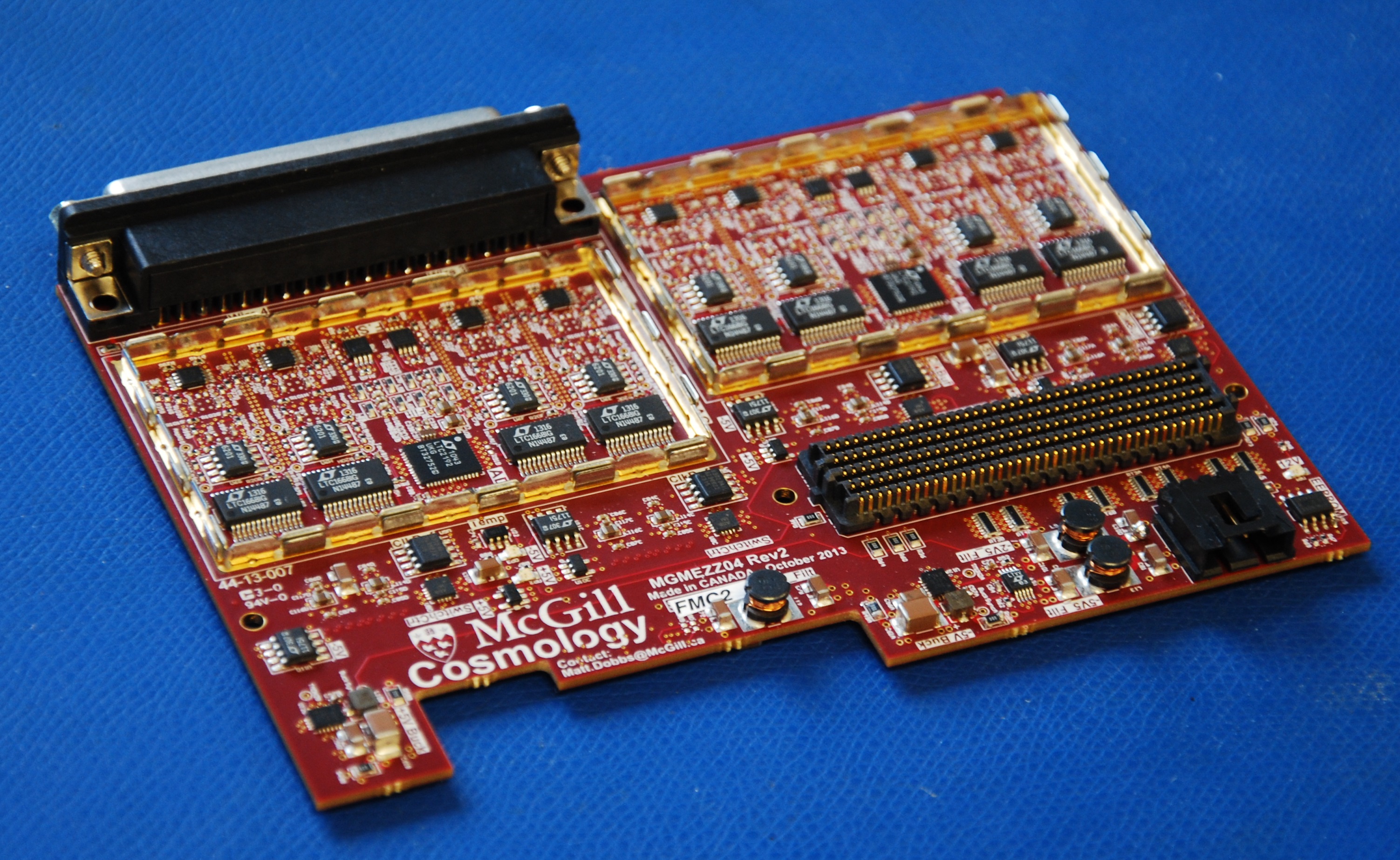}
\caption{
The ICE system \dfmux\ digitization/synthesis daughter board is shown.  Each
daughter board handles four multiplexing modules with a multiplexing factor of
up to 128 presently supported by these warm electronics. This means a single
crate of ICE motherboards can bias and readout 16,000 TES detectors.
Enhancements to the system firmware will likely allow the multiplexing factor
to increase to 256 detectors per pair of wires. 
New detector systems that are read out with frequency multiplexing such as Kinetic
Inductance Detectors (KIDs) are also under development for CMB experiments, the ICE
system can be specialized for these applications using higher sample-rate FMC
daughter boards that are available commercially, or by developing custom
daughter boards. }
\label{FigDfMuxMezz}
\end{figure}

For the \dfmux\ application, the ICE system uses custom FMC daughter boards
shown in Figure~\ref{FigDfMuxMezz}. Each ICE motherboard interfaces two
\dfmux\ daughter boards, together providing the readout for eight multiplexing
modules.

Each \dfmux\ daughter board multiplexing module includes two 16-bit
DACs operating at 20~MSPS, \hilight{providing the
detector bias and the nuller sinusoid combs}. The DACs are selected
for excellent low-frequency noise performance, since low-frequency noise is
modulated as a side-band on the bias carriers~\citep{2008ITNS...55...21D},
polluting the detector bandwidth. Amplification and anti-aliasing circuits
follow each DAC and a switched resistor network allows the gain to be adjusted
to 16 different levels. The error signals from the SQUID are amplified and
digitized at 20~MSPS with a 14~bit ADC. All ADCs and DACs in the array are
sampled synchronously. While a 20~MSPS sample rate is presently used, the
hardware itself supports sample rates up to 50~MSPS.  In addition to the FMC
daughter board, there is a specialized SQUID controller circuit board that
biases the SQUIDs and conditions the signals at the cryostat boundary. Several
cryogenic circuit boards house the multiplexer cold components, including the
SQUIDs and inductor-capacitor resonant circuits.

The custom firmware and digital signal processing techniques that were used
for the legacy \dfmux\ application are described in
\citet{2010arXiv1008.4587S}. Since then, a new DSP strategy that uses
polyphase filter banks (PFBs) and Fast Fourier Transforms (FFTs) to modulate
kHz-frequency sinusoids up to the required frequency, and demodulate sky signals
down from the carrier frequency to baseband, has been implemented. 
This new strategy makes more efficient
use of resources than the legacy firmware, allowing for higher multiplexing
factors, lower FPGA operating temperature, and higher sample rates.

For
\sptthreeg, with a multiplexing factor of 68, each crate provides readout for
8,700 detectors (there are two crates total for \sptthreeg) and draws less than
1~kW of power. The system is clocked with a single 10 MHz GPS-disciplined
reference and time-stamped with IRIG-B protocol signals provided by a GPS
receiver. For the \dfmux\ application, the demodulated and integrated data
rate is small. Each motherboard sends its data to the on-board ARM processor,
which forwards it through a gigabit Ethernet port using multicast UDP to the
experiment's control computer for storage and further software processing.

The ICE system \dfmux\ readout electronics presently supports multiplexing
factors of up to 128. The new PFB modulation and demodulation
strategy should allow a single ICE motherboard to process 8
readout modules each with a multiplexing factor of 256. In this configuration,
a single crate of motherboards will bias $\sim$32,000 detectors. An array of
ICE crates of the size being deployed for CHIME would provide readout for
250,000 detectors. However, present cryogenic circuits used for frequency
domain multiplexing support a maximum factor of 68---increased channel count
is an active area of technology development.

New detector systems that are read out with frequency multiplexing such as
KIDs~\citep[e.g., ][]{McHugh:2012dy} are also under
development for CMB experiments and hold great promise. The ICE system can be
specialized for these applications using higher frequency FMC daughter boards
that are available commercially, or by developing custom daughter boards. The
PFB modulation and demodulation can be used directly, by modulating the
carriers to higher frequency with appropriate FFT signal processing. For
observatories that plan to deploy both TES and KID focal planes, the ICE
system allows for the same core electronics to be used for the readout of both
detector systems.

\subsection {Radio Interferometry: The CHIME Example}
\label{subsec_chime}

CHIME was introduced in 
Section~\ref{sec_introduction}. It is a radio interferometer consisting of 
four 100~m $\times$ 20~m cylindrical dishes instrumented with 1024 dual-polarization 
feeds.  CHIME is designed to map the density of 
neutral hydrogen in the universe by measuring the redshifted 21-cm hydrogen 
emission line from 800~MHz to 400~MHz (redshift 0.8 to 2.5). By observing the 
apparent angular scale of the Baryon Acoustic Oscillations in large scale structure 
as a function of redshift, or distance,~\citep[see e.g.,\ ][Ch.\ 4]{2013PhR...530...87W} 
CHIME can directly observe the expansion history of the universe providing a new probe of dark
energy.
\hilight{A separate computer backend, `CHIME-FRB',} 
will allow it to trigger on high-dispersion measure
radio transients, probing the FRB phenomena~\citep[see e.g.,\
][]{2016MPLA...3130013K}.

CHIME employs a novel hybrid cylindrical design optimized for rapid, wide field
imaging. The cylinders are fixed with no moving parts, forming a transit
instrument with an instantaneous field of view of 90~degrees by 1-2 degrees
\hilight{and a synthesized beam resolution of 20-40 arcminutes}.
Modern low-noise amplifiers and the immense digital processing power provided
by the ICE system and a GPU X-engine removes the necessity for analog
beam-forming that has been common for large interferometers in the past.

The 2048 analog signals originating at the 1024 dual polarization feeds are
amplified on the focal line using low-cost, low-noise amplifiers that provide
a noise equivalent system temperature of less than 50~K across the band. The
signals are transmitted 60 meters on coaxial cables to nearby RF-shielded
enclosures. The signal is then amplified, band-pass filtered to 400-800 MHz,
and input to custom FMC digitizer daughter boards on the ICE motherboards.

The CHIME correlator is based on an FX design, \hilight{where 
the ICE system is used to implement the data acquisition and
Fourier transform channelization (F-Engine)}. The ICE system also
handles the majority of the corner-turn
networking~\citep{ICE_CORNER_TURN_2016}. A GPU-based X-engine was chosen to
compute the spatial correlation matrix providing signal processing
flexibility at the cost of larger power consumption. The flexibility afforded
by the GPU X-engine allows the calibration algorithms to be easily fine-tuned
as we better understand the instrument, and also allows easy
expansion, such as the introduction of additional parallel processing
pipelines that feed auxiliary pulsar and FRB backends.

\begin{figure} [htbp] \centering
\includegraphics[width=.8\textwidth]{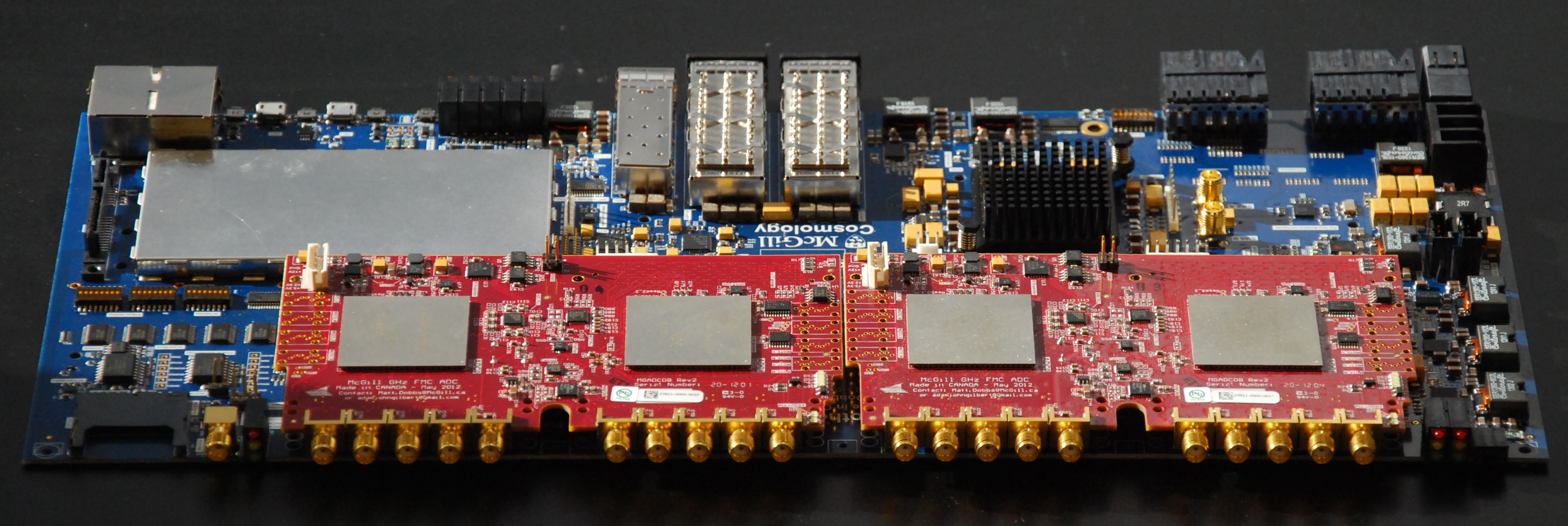}
\caption{ICE Motherboard (blue) is shown with two FMC daughter boards (red)
installed and the front-panel removed. Each daughter board provides eight
inputs capable of sampling up to 1.25 GSPS at 8~bits. For CHIME, they are
operated at 800~MSPS.}
\label{FigChimeMezzOnMB}
\end{figure}

Each ICE motherboard is equipped with two FMC digitizer daughter boards, shown
in Figure~\ref{FigChimeMezzOnMB}, which specialize the ICE motherboards for
the radio interferometry application. For CHIME, 128 motherboards are needed
to digitize the 2048 input signals. They are housed in eight crates equipped with ICE
backplanes, located in four racks distributed between two RF shielded rooms that
are housed inside two separate 20~ft shipping containers. Chilled water is
brought into the RF rooms for radiators and fans that blow cold air into each
crate \hilight{and efficiently extract heat from the rooms.}

Each digitizer daughter board has two EV8AQ160 
ADCs, which together provide eight inputs per daughter board, capable of
sampling up to 1.25 GSPS at 8~bits. For CHIME, they are operated at 800~MSPS
in the second Nyquist zone.

The EV8AQ160 converters have reduced channel modes which allow the sampling
rate to be increased to 2.5~GSPS (for two channels per ADC) and 5~GSPS (for a
single channel per ADC). The daughter board has its input matching tuned
to provide $>15$~dB return-loss performance in the 400-800~MHz frequency range
with a 2~dB insertion loss. This return loss bandwidth can be adjusted with
different matching components. The high-performance ADC supports analog inputs
with 500~mV peak to peak and provides 7.5 effective bits, a spur free
dynamic range of 52~dBc, and a signal to noise ratio of 48~dB.
The daughter boards have an on-board phase-locked-loop  (PLL)
to clock both ADCs and typically have a jitter $<0.4$~ps RMS. 
The PLL output
frequency and phase can be controlled  via an interface to the motherboard ARM
processor. An on-board EEPROM is present and is used to store calibration
parameters \hilight{and board metadata}.
The CHIME daughter board can interface to any FMC compliant FPGA based motherboard
that has fully populated HPC connector.

Within the FPGA on each ICE motherboard, the data from each of the 16
digitizer inputs is processed with custom firmware. It is packaged into 2048
sample frames and input to a PFB 
 that feeds a 2048-point
FFT, \hilight{both from the CASPER firmware library}. Complex gain corrections are applied to every
frequency bin, and the data is reduced to 1024 4-bit real + 4-bit imaginary
complex values per frame. The ICE system F-Engine outputs a total of 6.5~Tbit/s that
must be transmitted to the GPU X-engine.

At this point, all data from one antenna across the full bandwidth is in one
location within the FPGA F-engine. Each spatial correlation node (X-engine)
requires the data from all antenna feeds, but for a subset of the frequency
band.
This requires an exchange of data between ICE motherboards
called the corner-turn. The hardware and firmware developed for the ICE system
corner-turn is described in~\citet{ICE_CORNER_TURN_2016}. Briefly:

\begin{itemlist}
	\item The firmware within each ICE motherboard re-packetizes the data internally
	      to create 16 packets containing data from
	      16 analog inputs but 1/16th (64) of the frequency bins. The data is sent over the
          backplane high-speed serial links such that each FPGA ends up with the same set of frequency bins from all 16 boards.
	\item The data is re-packetized again. Each ICE motherboard sends half of
          its frequency bins to a motherboard in an adjacent ICE crate using four high-speed serial links in QSFP+ connectors located on the backplanes.
           Every ICE motherboard now has data from 512 analog inputs but 32 out of 2048 frequency bins.
	\item The packets are reordered again and split into eight streams containing data from 512
           analog inputs and 4 frequency bins. These streams are sent to eight different GPU nodes in the X-engine over eight 10~GbE
           links available through QSFP+ connectors located on each motherboard.
\end{itemlist}

Each of the 256 GPU Nodes has four 10~GbE ports and aggregates the data
directly from four ICE motherboards, each located in a different ICE crate.
Each GPU node then correlates and integrates data for four frequency bins from
all 2048 analog inputs. The integrated, low-rate data from all 256 GPU nodes
is forwarded to a data server over a 1~GbE link.
The GPU X-engine also performs real time beam forming  and forwards the data
to the FRB back-end. See \citet{Denman:2015ec,Recnik:2015ev,Klages:2015em} for
more information about the CHIME GPU X-engine.

A GPS receiver with a disciplined crystal oscillator and multiple outputs is
used to generate a 10 MHz reference clock signal and an IRIG-B time signal, both of which are
sent to each of the eight ICE backplane inputs.
\hilight{If needed, the 10 MHz clock can instead be
provided by a Hydrogen maser.} Synchronization of the array is performed by instructing all
boards to start ADC sample acquisition and framing on the rising edge of the
10 MHz clock following a specified IRIG-B timestamp. The delay of the
synchronization signal reaching the ADC chips can be adjusted by the FPGA to
ensure a reproducible starting point. Each data frame is tagged with a 48-bit frame
counter value, and the user can periodically capture the IRIG-B time
corresponding to the current frame in order to relate the frames with an absolute
time.

The construction of the CHIME telescope is nearing completion at the Dominion
Radio Astrophysical Observatory (DRAO) in Penticton, B.C., and the experiment is
entering its commissioning phase. When CHIME comes online this year, measured
in number of digitizer inputs squared times bandwidth, it will be the largest
radio correlator that has been built.

The CHIME Pathfinder (two 37~m~$\times$~20~m cylinders
instrumented with 128 dual-polarization feeds) is a
scaled down prototype version of CHIME  and has been in operation since 2015.
The Pathfinder is used to test the hardware (including the ICE system),
software, calibration techniques, and analysis strategy that will be used in
CHIME.
Operation of the Pathfinder and lab tests show that the data acquisition,
channelizing, and corner-turning are performing according to specifications and
without error.

The ICE system is also being deployed for the Hydrogen Intensity and Real-time Analysis 
eXperiment~\citep[HIRAX,][]{2016arXiv160702059N} in South Africa, which will consist of
1024 6-m dishes, operating with dual-polarization \hilight{feeds} from 400-800~MHz. HIRAX is
complementary to CHIME, having the same science goals and sharing technology,
but observing the Southern sky.

The ICE radio interferometry firmware has enough flexibility that it can be
configured as a single-board (16 input) correlator, single-crate (256 input)
correlator (e.g.,\ the CHIME Pathfinder), a dual-crate (512 inputs) correlator
(four of these are used for CHIME), and larger implementations.  For smaller
implementations, such as an eight channel correlator, an ICE motherboard can
be configured with firmware that also performs the spatial correlation, allowing for
an end-to-end FX correlator on a single board.

\hilight{This firmware flexibility also allows the ICE system to be used in applications such
as very-long-baseline interferometry (VLBI). The 
high-speed 8-bit digitization, followed by channelization and compression to 4 bits,
is well suited to VLBI, reducing the mixing of
RFI caused by traditional 1 and 2 bit quantizations. A single ICE motherboard has been deployed
at the Algonquin Radio Observatory (ARO) in Ontario, where the two polarizations of
a 46-m radio telescope are digitized and the F-engine output (channelized data) is
streamed directly to a disk array of twenty 8~TB disks to be correlated offline with the
respective channelized data from the CHIME pathfinder correlator.
With this setup, successful VLBI fringes have been detected on early experiments of
the Crab Pulsar (PSR B0531+21) over the 400-800 MHz band.}

The ICE system can be applied without hardware modifications for other radio
interferometry applications. The EV8AQ160 ADCs employed for the digitizer
daughter boards have an analog input bandwidth of 5~GHz, allowing systems to
be designed to alias sample other portions of the bandwidth from 0-5~GHz.
\citet{ICE_CORNER_TURN_2016} presents other configurations that allow the ICE
system to be scaled to much larger input count.

\section{Conclusion}
\label{sec_conclusions}

We have presented an overview of the ICE electronics system designed for
scientific instruments that require substantial digital signal processing
capabilities. With a reconfigurable FPGA, an on-board ARM processor,
compatibility with industry-standard mezzanines, and massive communications
bandwidth, the system is flexible and scalable and is well suited to
implement the data acquisition and large-scale signal processing of modern
telescopes.

The system has been adopted for several leading observatories in the radio
and millimeter band. Since the system is designed for scaling from small
benchtop applications to large deployments with dozens or more boards, it
has also been used at a large number of universities and institutions
for technology research and development. Hundreds of ICE motherboards have
been built to date and have been distributed for operation at more than
eighteen universities or institutions on five continents.

\section{Acknowledgements}

We have benefited enormously from the close collaboration within the SPT,
POLARBEAR, Simons Array, CHIME, and HIRAX collaborations. We are
grateful to the National Research Council Canada which operates the Dominion Radio
Astrophysical Observatory, the U.S.\ National Science Foundation which
operates the Amundsen-Scott research station, and the Chilean CONICYT Atacama
Astronomical Park that hosts the Simons Array and POLARBEAR.

We acknowledge generous support from Xilinx University Programs which enables
both our prototyping and the training we provide for students within our
laboratories. We acknowledge funding from the Natural Sciences and Engineering
Research Council of Canada, Canadian Institute for Advanced Research, Canadian Space Agency,
Canadian Foundation for Innovation and le Cofinancement gouvernement du Qu\'ebec-FCI.
\hilight{Work at Argonne National Lab is supported by UChicago Argonne, LLC, Operator of 
Argonne National Laboratory (Argonne). Argonne, a U.S. Department of Energy 
Office of Science Laboratory, is operated under Contract No. DE-AC02-06CH11357.}



\bibliographystyle{ws-jai}
\bibliography{references_arxiv}   

\end{document}